\documentstyle[12pt]{article}
\newcommand{\e}{\; {\rm e} }

\newcommand{\be}{\begin{eqnarray} }
\newcommand{\ee}{\end{eqnarray} }
\newcommand{\beq}{\begin{equation} }
\newcommand{\eeq}{\end{equation} }

\begin{document}
\begin{center}
{\Large \bf
Topological charge and topological susceptibility in connection with
translation and gauge invariance
}
\end{center}
\vskip 1.5cm
\begin{center}
{ \large A.M.~Kotzinian}
\footnote{E-mail address: Aram.Kotzinian@cern.ch},
{ \large O.Yu.~Shevchenko}
\footnote{E-mail address: shevch@nusun.jinr.ru} and 
{ \large A.N.~Sissakian}
\footnote{E-mail address: sisakian@jinr.ru}\\ 
\vspace{1cm}
{\it Joint Institute for Nuclear Research\\
Dubna, Moscow region 141980, Russia}\\

\end{center}
\vskip 1.5cm
\begin{abstract}
It is shown that the evaluation of the expectation value (EV) of topological charge density 
over $\theta$-vacuum is reduced to investigation of the Chern-Simons term EV.
An equation for this quantity is established and solved. 
EV of the topological charge density at an arbitrary $\theta$ occurs  
equal to zero and, as a consequence, topological susceptibility 
of both QCD and pure Yang-Mills vacua defined in a Wick sense is equal to zero, 
whereas when defined in a Dyson sense it differs from zero by the quantity
proportional to the respective condensate of the chromomagnetic field. 
Thus, the usual Witten-Veneziano formula for the $\eta^{'}$ meson mass is modified.
\end{abstract}
\newpage

The effects connected with the nontrivial topological
configurations of the gauge fields attract a great attention
in modern physics. In this respect the QCD topological susceptibility
\be
\chi_{QCD} =\int d^4x \langle Tq(x)q(0) \rangle
\ee
is the quantity of a special importance because it enters
as a key object in a lot of physical tasks, in particular,
in such important puzzles as a famous $U(1)$ problem [1-5]
(see [6] for a recent review) and the "spin crisis" [7].
In Eq.(1) $q(x)$ is the topological charge density
\be
q(x)=\frac{g^2}{32\pi^2}{\rm tr} F^a_{\mu\nu}(x)\tilde{F}^{\mu\nu}_a(x)
\ee
related with the Chern-Simons current, $K_\mu(x)$ by
\be
q(x)=\partial^\mu K_\mu(x),
\ee
where
\be
K_\mu=\frac{g^2}{32\pi^2}\epsilon^{\mu\nu\rho\sigma}
A^a_\nu\left(F^a_{\rho\sigma}
-\frac{g}{3}f_{abc}A^b_{\rho}A^c_{\sigma}\right).
\ee 

It is well known, that topological  susceptibility $\chi_{QCD}$ is equal
to zero in all orders of perturbation theory and, also,
that this quantity is just zero in the presence of even
one massless quark (Crewther theorem [2]). 

In this paper the consideration 
based on the fundamental translation and gauge
symmetries  will be performed which will allow to draw some
unexpected conclusions about the  topological charge and
susceptibility.

Let us proof the following {\bf statement}.

EV of the topological charge density
$(\theta|q(0)|\theta)=(1/VT)(\theta|Q|\theta)$
over $\theta$-vacuum
with an arbitrary $\theta$ is equal to zero if 
EV of operator $K_\mu(0)$ over  $\theta$ -vacuum
exists, i.e.,
\be
|(\theta|K_\mu(0)|\theta)|<\infty,
\ee
where symbol $|\theta)$ denotes the $\theta$-vacuum state normalized to unity:
\be
(\theta|\theta)=1.
\ee

This statement directly follows from translation invariance of 
$\theta$-vacuum: 
\be
&&(\theta|q(0)|\theta)=(\theta|\partial^\mu K_\mu(0)|\theta)\nonumber\\
&&=
-i(\theta|[\hat{P}^\mu,K_\mu(0)]|\theta)
=-i(P_{\theta}^{\mu}-P_{\theta}^{\mu})(\theta|K_\mu(0)|\theta)=0.
\ee

 The key point here is the condition (5) which, as we will see below,
in $A_0=0$ gauge is equivalent to the condition
\be
\label{ksdjhf}
|(\theta|W_{CS}(0)|\theta)|<\infty,
\ee
where\footnote{
In the literature Chern-Simons term $W_{CS}[A]$ is also often called
"collective coordinate" (see, for example, [5]) and is denoted by $X[A]$.} 
\be
W_{CS}(t)\equiv\int d^3x K_0(x)
=
\frac{g^2}{32\pi^2}\int d^3x\left(\epsilon^{ijk}A^a_i\left(F^a_{jk}
-\frac{g}{3}f_{abc}A^b_{j}A^c_{k}\right)\right)
\ee
is the  Chern-Simons operator (see [8] for review).
However, as we will see, within the conventional formulation of
$\theta$-vacuum theory rather
amazing situation arises. On the one hand the condition (\ref
{ksdjhf}) {\it is not satisfied} due to the gauge
non-invariance of the operator $W_{CS}$ with respect to the "large",
topologically nontrivial gauge transformations. Nevertheless, despite EV 
$\langle\theta^{'}|W_{CS}(0)|\theta\rangle $ is more singular function
than $\delta(\theta^{'}-\theta)$ at $\theta^{'}\to \theta$
(namely, it behaves as $\delta^{'}(\theta^{'}-\theta)$ in this limit),
the EV of the topological charge density
over $\theta$-vacuum is just zero again. 

Since we deal with the gauge invariant quantity (EV of the topological
charge) let us choose the Weil gauge 
\be
A_0=0, 
\ee
which allows to essentially simplify a consideration. 
Choosing the 
periodic boundary conditions in the space directions
(topology of a hypercylinder oriented along the time axis)
one has
\be
\int d^3x\ \partial^iK_i(t,\vec x)=0,
\ee
and the expression for the topological charge 
\be
Q\equiv\int d^4xq(x)=\int d^4x\partial^\mu K_\mu
\ee
becomes
(see, for example, [8,9])
$$Q=W_{CS}(t=\infty)-W_{CS}(t=-\infty).$$

EV  $(\theta|\hat O|\theta)$ of an arbitrary
operator $\hat O$ over $\theta$-vacuum is defined as (see, for example [8,9])
\be
(\theta|\hat O|\theta)=\frac{\langle\theta|\hat O|\theta\rangle}
{\langle\theta|\theta\rangle}
\ee
where $|\theta\rangle$
is, simultaneously, the eigenfunction of the full $QCD$ Hamiltonian
$H$ and of the unitary operator ${\rm T}_\nu$ of the large gauge 
transformations with a winding number $\nu$:
\be
H|\theta\rangle=E_\theta|\theta\rangle,\\
T_\nu|\theta\rangle={\rm e}^{-i\theta\nu}|\theta\rangle,
\ee
i.e, the state $|\theta\rangle$ is, up to a phase multiplier, gauge invariant 
against the large gauge transformations. 
Notice also that on the contrary to (6) the states $|\theta\rangle$ are
normalized as
\be
\langle\theta'|\theta\rangle=\delta(\theta'-\theta).
\ee
so that the prescription (13) reads
\be
\langle\theta'|\hat O|\theta\rangle=(\theta|\hat O|\theta)\delta(\theta'
-\theta),
\ee
i.e., $(\theta|\hat O|\theta)$ is just the limit at $\theta'\rightarrow\theta$
of the multiplier at $\delta$-function in the expression for
$\langle\theta^{'}|\hat O|\theta\rangle$.

Since we are interested in the quantity 
\be
(\theta|q(0)|\theta)=(VT)^{-1}(\theta|Q|\theta),
\ee
we will keep the normalization factor $(VT)^{-1}$.
Using (11) and the Heisenberg equations one easily gets 
\be
&&(VT)^{-1}\langle\theta'|Q|\theta\rangle=
(VT)^{-1}\int dt{\rm e}^{i(E_{\theta'}-E_\theta)t}\langle\theta'|
\int d^3xq(0,\vec x)|\theta\rangle
\nonumber\\
&&=2\pi(VT)^{-1}\delta(E_{\theta'}-E_\theta)\langle\theta'|
\dot W_{CS}(0)|\theta'\rangle
\nonumber\\
&&=2\pi(VT)^{-1}\delta(E_{\theta'}-E_\theta)\langle\theta'|
-i[W_{CS}(0),H]|\theta\rangle
\nonumber\\
&&=2\pi i(VT)^{-1}\delta(E_{\theta'}-E_\theta)
(E_{\theta'}-E_\theta)
\langle\theta'|W_{CS}(0)|\theta\rangle,
\ee
and, thus, the task now is to evaluate EV 
$\langle\theta'|W_{CS}(0)|\theta\rangle$.

The remarkable  property of the Chern-Simons term 
is its transformation law under  
topologically nontrivial (often called "large" [8]) gauge transformations
\be
A_i\rightarrow A_i^{\Omega_{\nu}}=\Omega_{\nu}A_i\Omega_{\nu}^{-1}
+\partial_i\Omega_{\nu}\Omega_{\nu}^{-1}\ 
\ (i=1,2,3; \,\,\,\Omega=\Omega(\vec x) ),
\ee
with topological index (winding number) $\nu$.
Namely, the quantity $W_{CS}[A]$ is not gauge invariant 
under (20) but  
transforms as 
\be
\label{aaa}
W_{CS}[A]\rightarrow W_{CS}[A^{\Omega_{\nu}}]=W_{CS}[A]+\nu,
\ee
i.e. it only shifts by the winding number $\nu$ of the respective gauge
transformation. 

The compatibility of the quantum
\be
W_{CS}[A^{\Omega_{\nu}}]=T_\nu W_{CS}[A]T_\nu^+=W_{CS}(A)+[T_\nu,W_{CS}(A)]
T_\nu^{-1}
\ee
and classical (21)
gauge transformation laws of the Chern-Simons term (9) gives
rise to the commutation law
\footnote{Here one uses         
that 
$[T_\nu,H]=0$.}
\be
[T_\nu,W_{CS}(t)]=[T_\nu,W_{CS}(0)]=\nu T_\nu.
\ee

Now one already can evaluate
$\langle\theta{'}|W_{CS}(0)|\theta\rangle$. 
Indeed, due to the unitarity of the operator $T_\nu$ and Eq. (15) one has
\be
\langle\theta{'}|[T_{\nu},W_{CS}(0)]|\theta\rangle=\left(\e^{-i\nu{\theta}'}-\e^{-i\nu\theta}
\right) 
\langle\theta{'}|W_{CS}(0)|\theta\rangle.
\ee
On the other hand the commutation law (23) together with Eqs. (15), (16)
give
\be
\langle\theta{'}|[T_\nu,W_{CS}(0)]|\theta\rangle=
\nu{\rm e}^{-i\nu\theta}\delta(\theta-\theta{'}).
\ee
Comparing (24) and (25) one gets the basic for what follows equation
\be
\frac{1}{\nu}\left(\e^{-i\nu(\theta{'}- \theta)}-1\right)
\langle\theta{'}|W_{CS}(0)|\theta\rangle =
\delta(\theta{'}-\theta).
\ee 
So, one has to solve the equation
\be
\frac{1}{\nu}\left(\e^{-i\nu z}-1\right)
f(z,\theta) =
\delta_{2\pi}(z),
\ee
where $z\equiv \theta{'}- \theta$, 
$f(z,\theta)\equiv \langle\theta{'}|W_{CS}(0)|\theta\rangle$,
and $\delta_{2\pi}(z)\equiv \delta(\theta{'}-\theta)$ 
is $2\pi-$periodic $\delta-$function.
Expanding  
$f(z,\theta)$ and $\delta_{2\pi}(z)$
in the Fourier series
\be
f(z,\theta)=\frac{1}{2\pi}\sum_{n=-\infty}^{\infty}\tilde f_n(\theta)\e^{inz}, \quad
\delta_{2\pi}(z)=\frac{1}{2\pi}\sum_{n=-\infty}^{\infty}
\e^{inz},
\ee
one easily obtains instead of (27) the difference equation for Fourier image $\tilde f$
\be
\tilde f_{n+\nu}(\theta)-
\tilde f_n(\theta)
 =\nu
\ee
with the solution
\be
\tilde f_n(\theta)= n + C(\theta),
\ee
where $C(\theta)$ is some arbitrary function of $\theta$.

Thus, the the general solution 
\footnote{It is of importance that (31) is the solution of (26) only
  if the winding number $\nu$ is an arbitrary {\it integer} number
  (the existence of the fractional winding numbers was advocated
  in~\cite{2}). 
Otherwise, one can not perform the necessary change $n-\nu \to n$ in the sum
$\sum n \e^{i(n-\nu)z}$.}
of equation (25) reads
\footnote{Notice that for any 
{\it gauge invariant} 
operator 
$[T_{\nu}, O_{g.inv}]=0$ and, therefore, Eq. (31) is replaced by
$
[\exp(-i\nu\theta{'})-\exp(-i\nu\theta)] \langle\theta^{'}|O_{g.inv} |\theta\rangle=0
$ with the general solution
$
\langle\theta^{'}|O_{g.inv} |\theta\rangle=C(\theta)\delta(\theta^{'}-\theta).
$
}  
\be
\langle\theta^{'}| W_{CS}(0)|\theta\rangle = - i\delta{'}(\theta^{'}-\theta)+ 
C(\theta)\delta(\theta^{'}-\theta),
\ee
where 
$\delta{'}(\theta^{'}-\theta)=(i/2\pi)\sum n \e^{inz}$.

Since
\be
\lim_{\theta^{'}\to \theta}\left(\frac{2\pi}{T}
\delta\left(E_{\theta^{'}}-E_{\theta}\right)\right)\left(E_{\theta^{'}}-E_{\theta}\right)
=\left(E_{\theta}-E_{\theta}\right)=0,
\ee
the term $C(\theta)\delta(\theta^{'}-\theta)$ in the solution (31) does not 
contribute to the coefficient at
$\delta(\theta^{'}-\theta)$
in the r.h.s. of (19) and, thus
\be
(VT)^{-1}\langle\theta{'}|Q|\theta\rangle = 2\pi(VT)^{-1}
\delta^{'}(\theta^{'}-\theta)
\left[\left(E_{\theta'}- E_{\theta}\right)\delta\left(E_{\theta'}- 
E_{\theta}\right)\right].
\ee
Considering $\theta^{'}$ as a variable whereas $\theta$ is kept fixed, one gets
\be
(VT)^{-1}\langle\theta{'}|Q|\theta\rangle &=& 
-2\pi(VT)^{-1}\delta(\theta^{'}-\theta) 
\frac{\partial E(\theta^{'})}{\partial \theta^{'}}\nonumber \\
&\times&
\Bigl[\delta^{'}\left(E_{\theta'}- E_{\theta}\right)
\left(E_{\theta'}- E_{\theta}\right) +
\delta\left(E_{\theta'}- E_{\theta}\right)\Bigl],
\ee
On the first sight the expression in the square brackets 
is equal to zero since usually $x\delta^{'}(x)=-\delta(x)$.
However, this is not correct conclusion and one has to properly work with the
generalized function
$\left(E_{\theta'}- E_{\theta}\right)\delta^{'}\left(E_{\theta'}- E_{\theta}\right)$
when takes the limit $\theta^{'}\to \theta$, i.e., 
$\left(E_{\theta'}- E_{\theta}\right) \to 0 $. 
Indeed, let us consider the generalized function 
\be
\Delta (x)\equiv x\delta^{'}(x).  
\ee
Then
\be
\Delta (0)&=& \int\limits_{-\infty}^{\infty}dx \delta(x)\Delta(x) \nonumber\\
&\equiv& \int\limits_{-\infty}^{\infty} dx \delta(x)[ x\delta^{'}(x)]
=-\int\limits_{\infty}^{\infty} dx\delta(x)[ \delta(x)+x\delta^{'}(x)] \nonumber \\
&=& -\int\limits_{\infty}^{\infty}dx \delta(x)\Delta(x)- 
\int\limits_{\infty}^{\infty} dx\delta(x)\delta(x) 
= -\Delta(0)- \delta(0), \nonumber
\ee
and, thus\footnote{The quantity $\Delta(x)$ is equal to $-\delta(x)$ as a
generalized function only in the convolution with a function $F(x)$ 
satisfying the condition $ xF^{'}(x)|_{x=0}=0$.
However, this is just not the case for the choice
$F(x) = \delta(x)$. }
\be
\Delta (0)\equiv [x\delta^{'}(x)]\Bigl |_{x=0} =-\frac{1}{2}\delta (0) .
\ee
In particular,
\be
\lim_{\theta^{'}\to \theta}\left[\left(E_{\theta^{'}}-E_{\theta}\right) 
\delta^{'}\left(E_{\theta^{'}}-E_{\theta}\right)\right]  
=-\frac{1}{2}\delta\left(E=0 \right)   
=-\frac{1}{2}\frac{T}{2\pi}. 
\ee

In accordance with the general prescription (17) and Eqs. (18), 
(34), (37)
one obtains
\be
(\theta|q(0)|\theta)=-\frac{\partial E_{\theta}}{\partial \theta}
\frac{2\pi}{VT}\left[-\frac{1}{2}\frac{T}{2\pi}+ \frac{T}{2\pi}\right]
= - \frac{1}{2}\frac{\partial \epsilon(\theta)}{\partial \theta}, 
\ee
where
\be
\epsilon(\theta) \equiv E_{\theta}/V
\ee
is the energy density of $\theta$-vacuum.

On the other hand, it is well known that the energy density $\epsilon(\theta)$
is  defined via the functional integral as (see, for example, [5,9])
\be
\epsilon(\theta)=i(VT)^{-1}{\rm ln}W_\theta,
\ee
where
\be
W(\theta)\equiv \int{\cal D}A{\cal D}[\bar \psi, 
\psi]\exp i\left[S_{QCD}+\theta Q\right].
\ee
In this picture one has 
%
\be
(\theta|q(0)|\theta)  
=(VT)^{-1}(\theta|Q|\theta)=
-i(VT)^{-1}\frac{\partial W[\theta]}{\partial\theta}
W^{-1}[\theta]
=-\frac{\partial\epsilon(\theta)}{\partial\theta},
\ee
whereas the second derivative of $\epsilon(\theta)$ with respect to $\theta$
just produces the {\it topological susceptibility} - the connected part
of the two-point correlator of the topological charge densities at 
zero momentum:
\be
\chi_\theta
=\int d^4x (\theta|{\rm T}q(x)q(0)|\theta)_{conn} 
=-(VT)^{-1}\frac{\partial^2{\rm ln}W_\theta}{\partial\theta^2}=i
\frac{\partial^2\epsilon(\theta)}{\partial\theta^2},
\ee
and
\be
\chi_{QCD}=\chi_{0}.
\ee

It is easy to see now that the only way to reconcile  Eqs. (38) and (42)  
is to put
\be
(\theta|q(0)|\theta)=-\frac{\partial\epsilon(\theta)}{\partial\theta} = 0.
\ee

Then one can see that the topological susceptibility defined by (43)
is also equal to zero
\be
\chi_{\theta}=\chi_0=0.
\ee

Let us now attempt to realize obtained result. 
On the first glance Eq. (46) is in a severe contradiction
with a standard point of view [2-6] that the quantity $\chi$
must differ from zero because it is directly connected 
with the solution of $U(1)$-problem and   
mass of $\eta'$-meson  
is explicitly
expressed via topological susceptibility. 
%
%
%
%
However, the situation perhaps is not so bad. 
The point is that there exist {\it two different topological
susceptibilities} in accordance with the different sense
of the time-ordering operation in the respective correlators. 

 Let us remind that the functional integral
representation (40)-(42) for a correlator means that 
the respective ${\rm T}$-product
in this correlator must be realized as the Wick (${\rm T}_W$)
time-ordering operation,
in which
all the derivatives are applied after the calculation of the field
convolutions (see [5] for the excellent review on this question).
On the contrary, the Dyson ${\rm T}$-ordering of two arbitrary
operators $A(x)$ and $B(y)$ (does not matter composite or not)
simply looks as
\be
{\rm T_D}[A(x)B(y)]=\theta(x_0-y_0)A(x)B(y)+\theta(y_0-x_0)B(y)A(x).
\ee

So, actually Eq. (46) have to be read as
\be
\chi^W_\theta=\chi^W_0=0,
\ee
where
\be
\chi_\theta^W\equiv
\int d^4x[(\theta|{\rm T}_W[q(x)q(0)]|\theta)_{conn},
\ee
but {\it it does not mean at all} that the 
Dyson topological susceptibility 
\be
\chi_{\theta}^D =\int d^4x[(\theta|{\rm T}_Dq(x)q(0)|\theta)-
{(\theta|q(0)|\theta)}^2]
=\int d^4x (\theta|{\rm T}_Dq(x)q(0)|\theta)_{conn}
\ee
is also equal to zero.

Indeed, the connection between Wick and Dyson susceptibilities
was found in [5] using the stationary perturbation theory 
in powers of $\theta$ with a result 
\be
\chi_0^W =i\frac{\partial^2 
\epsilon}{\partial\theta^2}\Bigl|_{\theta=0}
=\chi_0^D +i(0|\left(\frac{g^2}{8\pi^2}\vec B_a\right)^2|0),  
\ee                                                               
i.e., these susceptibilities differ by the  condensate
of the chromomagnetic field $\vec B_a$.
Thus, even despite that in accordance with (48) the quantity $\chi^W_0 $
equal to zero, the Dyson topological susceptibility $\chi_D^0$ 
can differ from zero by the nonzero value of the chromomagnetic condensate: 
\be
\chi_0^D=-i(0|\left(\frac{g^2}{8\pi^2}\vec B_a\right)^2|0).
\ee

On the other hand it is well known that namely Dyson ${\rm T}$-product (not Wick!)
can be represented as the sum over intermediate states with a result
\be
\int d^4x \e^{ik\cdot x}(0|{\rm T_D}(A(x)B(0))|0)_{conn}=
\sum_n(0|A(0)|n(\vec k))( n(\vec k)|B(0)|0)
\frac{i}{k^2-m^2_n}
\nonumber\\
+\mbox{\rm two particle contributions}\quad (k^2=k_0^2-\vec k^2).
\ee
As it was shown by Witten [3] (see [11], sect 5.1, for review)  
only the one particle contributions
survive in the sum over intermediate states in the large
$N_c$ limit. So,
\be
&&\chi_0^D{\bigl |}_{QCD}=\int d^4x\ 
(0|{\rm T_D}q(x)q(0)|0)_{conn}{\bigl |}_{QCD}
\nonumber\\
&&=\lim_{k\rightarrow0}\left[\sum_{n=mesons}|(0|q(0)|n(\vec k))|^2
\frac{i}{k^2-m^2}+\sum_{l=glubols}|(0|q(0)|l(\vec k))|^2
\frac{i}{k^2-m^2}\right]
\nonumber\\
&&=\chi_0^D{\bigl |}_{YM}+\sum_{n=mesons}
|(0|q(0)|n)|^2\frac{-i}{m_n^2},
\ee
and, in the leading order of $\chi PT$ where [6]
\be
(0|q(0)|\eta')=
(0|\frac{1}{2N_f}\partial^{\mu}J_{\mu}^5|\eta')=
\frac{1}{2N_f}f_{\eta'}m^2_{\eta'}
\ee
one gets instead of Witten-Veneziano [3,4] formula
\be
&&m^2_{\eta'}=-i\frac{4N_f^2}{f^2_{\eta'}}(\chi^D_0{\bigl |}_{YM}-
\chi^D_0{\bigl |}_{QCD})
\nonumber\\
&&=\frac{4N_f^2}{f^2_{\eta'}}
\left[
(0|\left(\frac{g^2}{8\pi^2}\vec B_a\right)^2|0){\bigl |}_{QCD}-
(0|\left(\frac{g^2}{8\pi^2}\vec B_a\right)^2|0){\bigl |}_{YM}\right].
\ee
It is of importance and seems to be a serious argument
in support of our model independent  consideration 
(based on the general principles of
translational and gauge invariance) that very similar
results  concerning
$ \eta^{'}-$meson mass   
was obtained within the different QCD inspired models.
These are Cheshire cat principle model [12] 
and, 
also,
squeezed gluon vacuum [13] 
and monopole vacuum [14] 
models
(compare\footnote{Comparing these formulas one has to use that
$\alpha_s=g^2/4\pi$ and $f_{\eta^{'}}=\sqrt{2N_f}f_{\pi}$.}
Eq. (56) with Eq. (14) of ref. [12] and, especially, with Eqs. (18) and (326) 
of
refs. [13] and [14], respectively).

Thus, we get rather unexpected result: Wick topological susceptibility
is equal to zero whereas the Dyson topological
susceptibility just proportional to the chromomagnetic condensate. 
The last circumstance allows one to get the mass formula (56)
for the $\eta'$-meson which directly express its mass via the
difference of the respective chromomagnetic condensates with and without quark inclusion.


\vskip0.3cm
\begin{center}
{\bf Acknowledgment}
\end{center}
The authors are grateful to E. Kuraev, S. Nedelko, V. Pervushin, I. Solovtsov, 
O. Teryaev and G. Veneziano for fruitful discussions.

\end{document}